\documentclass{Interspeech}

\interspeechcameraready

\title{Training-Free Voice Conversion with Factorized Optimal Transport}

\author[affiliation={1}]{Alexander}{Lobashev}
\author[affiliation={2}]{Assel}{Yermekova}
\author[affiliation={3}]{Maria}{Larchenko}

\affiliation{}{Glam AI}{San Francisco, US}
\affiliation{}{Independent researcher}{Astana, Kazakhstan}
\affiliation{}{Magicly AI}{Dubai, UAE}
\email{lobashevalexander@gmail.com, allessyer@gmail.com, mariia.larchenko@gmail.com}


\usepackage{comment}

\begin{document}

\maketitle

\begin{abstract}
This paper introduces Factorized MKL-VC, a training-free modification for kNN-VC pipeline.
In contrast with original pipeline, our algorithm performs high quality any-to-any cross-lingual voice conversion with only 5 second of reference audio. MKL-VC replaces kNN regression with a factorized optimal transport map in WavLM embedding subspaces, derived from Monge-Kantorovich Linear solution.  Factorization addresses non-uniform variance across dimensions, ensuring effective feature transformation. Experiments on LibriSpeech and FLEURS datasets show MKL-VC significantly improves content preservation and robustness with short reference audio, outperforming kNN-VC. MKL-VC achieves performance comparable to FACodec, especially in cross-lingual voice conversion domain.
\end{abstract}
    
    
    

\section{Introduction}
Any-to-any voice conversion (VC) aims to change the voice identity of a speaker to match a reference voice without training a specific model for each speaker-reference pair \cite{sisman2020overview}. The linguistic content of the speaker's utterance should remain unchanged, so the conversion targets only non-linguistic features (prosody) such as intonation, pitch, and timbre.

A recent study by Baas et al. (2023) has introduced kNN-VC, a simple and efficient pipeline for voice conversion based on k-nearest neighbors regression \cite{baas23_interspeech}. Its key idea is to replace an embedding in the source sequence with the closest embeddings taken from the target sequence and then decode the resulting sequence back into waveform output. The kNN-VC algorithm shows a state-of-the-art level with relatively long (5 minutes) reference audio recordings.

However, it fails to keep the same quality in the range of 1 min references: It is hard to find sufficiently similar sounds to cover the whole source utterance. With longer recordings, the quality is improved by the computational cost of the nearest neighbors search. Another limitation of kNN-VC is cross-lingual conversion, as it tends to introduce language-specific pronunciation from a reference audio into a source utterance. For example, the phoneme /r/ could be converted from the rolled /r/ into the guttural one, which may not be a desirable feature.

Note that the nearest-neighbor search in kNN-VC relies on an important property of an encoder's latent space: a meaningful cosine distance between embeddings. It exploits the observation that perceptually close sounds are encoded close to each other in the latent space. This may not be true for any model, but models trained in a self-supervised way with contrastive learning (as WavLM-Large \cite{chen2022wavlm}) indeed have this property.

Given that, one can justify the use of optimal transport theory. The main goal of optimal transport is to transform the source probability density to the target one with minimal cost, and the cost is assumed to be a distance.

\begin{figure}[t]
  \centering
  \includegraphics[width=\linewidth]{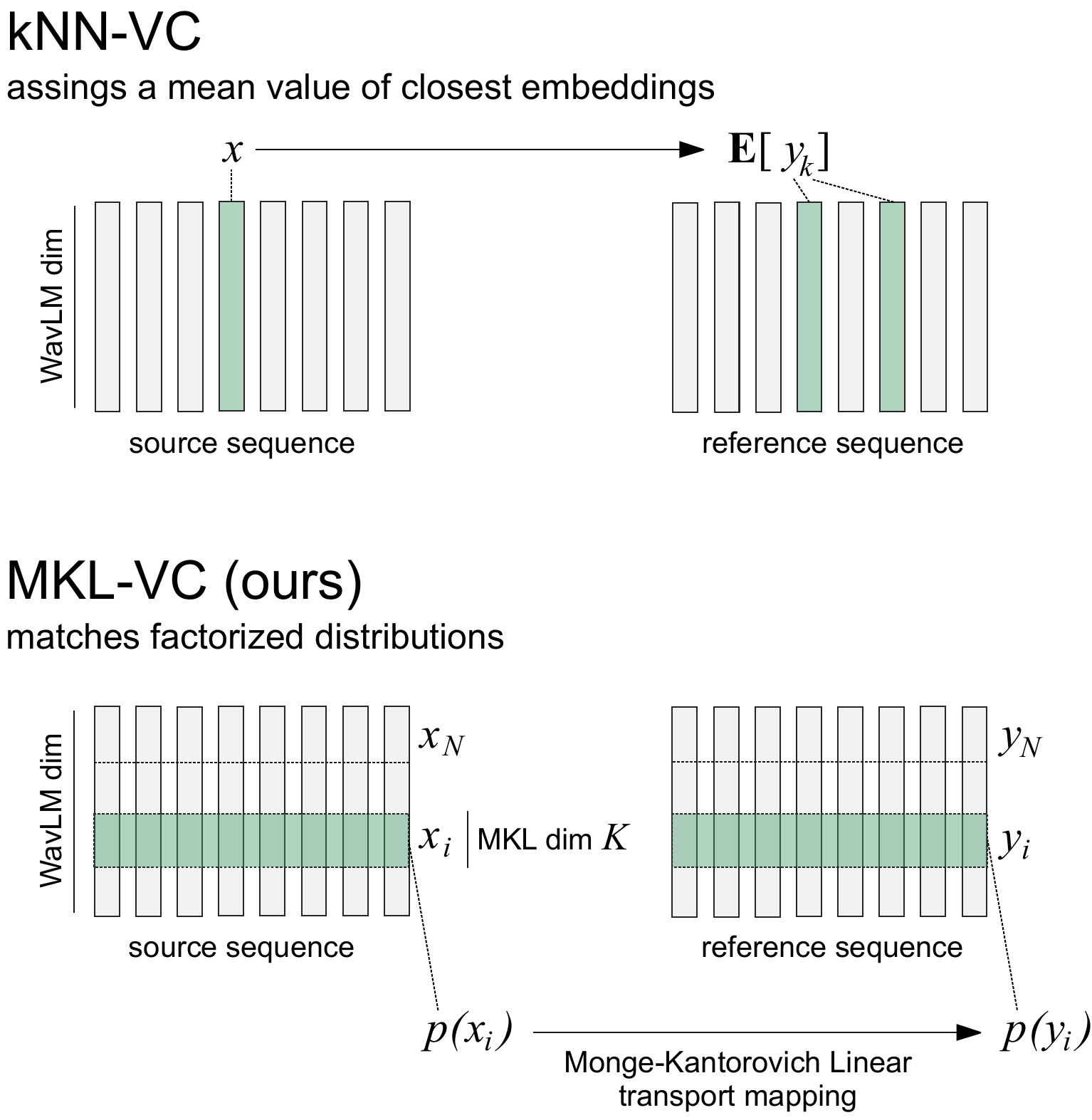}
  \caption{Our Factorized MKL method performs distribution matching per sorted dimensions of WavLM encoder.}
  \label{fig:scheme}
\end{figure}

Here we suggest to modify the kNN-VC pipeline by replacing kNN regression with a linear optimal transport map, as shown in Fig.\ref{fig:scheme}.
This map is defined by the explicit formula for optimal transport between multivariate Gaussian distributions, minimizing the quadratic transportation cost.
The contribution of this paper can be summarized as follows
\begin{itemize}
    \item We propose a training-free modification of the kNN-VC pipeline for any-to-any voice conversion based on the optimal transport theory \footnote{The code is available at https://github.com/alobashev/mkl-vc}.
    \item We demonstrate that this approach is superior in terms of content preservation and is robust with less than 1 min of reference audio.
    \item We argue that our technique is particularly beneficial for cross-lingual conversion and low-resource languages.
\end{itemize}

\section{Related Work}

Despite advances in the field of voice conversion, any-to-any algorithms are not that common \cite{walczyna2023overview}. The most recent ones include FACodec \cite{ju2024naturalspeech}, kNN-VC \cite{baas23_interspeech}, Seed-VC \footnote{\url{https://github.com/Plachtaa/seed-vc}}, FreeVC \cite{li2023freevc}, YourTTS \cite{casanova2022yourtts}, Diff-VC \cite{popov2022diffusion}, and VQMIVC \cite{wang2021vqmivc}.

The standard scheme for voice conversion is ``analysis-mapping-reconstruction'' pipeline \cite{sisman2020overview}, also described as ``encoder-converter-vocoder'' \cite{baas23_interspeech}\cite{walczyna2023overview}. The mapping stage considers different representations of the source and target speech and, in principle, can operate directly in the time domain of the input signal (PSOLA \cite{schnell2000synthesizing}). Mel-spectrogram is another popular choice, but recent studies have shifted towards a latent representation of a speech taken from a sound encoder. The most prominent examples of such encoders (also called codecs) are HuBERT \cite{hsu2021hubert}, WavLM\cite{chen2022wavlm} and FACodec\cite{ju2024naturalspeech}.

The mapping algorithms are very diverse between VC systems. The biggest challenge of this stage is to separate the speaker identity from the speech content and then perform artifact-free conversion. In the case of disentangled speech representation one can just replace the ``identity part'' in the latent vectors sequence (as the FACodec \cite{ju2024naturalspeech} baseline). Despite FAcodec's effectiveness in disentangling content and prosody representations, it is not universally perfect for all languages due to variations in linguistic structures, phonetic complexity, and prosodic patterns across languages, which can challenge its generalization capabilities.

\textbf{Optimal Transport in Voice Conversion}
Four OT-based algorithms (NOT, XNOT, SinkVC and FMVC) were proposed in the study by Asadulaev et al. (2024) \footnote{\url{https://arxiv.org/abs/2411.02402}}, but to our best knowledge only SinkVC can be used as a true any-to-any converter. FMVC seems to be not conditioned on unseen speakers and was reportedly tested on the ``unseen 100 utterance for the given speaker''.

SinkVC algorithm modifies the kNN-VC pipeline by replacing the kNN search with the Sinkhorn algorithm \cite{cuturi2013sinkhorn}, which solves the entropy-regularized optimal transport problem. Due to the simplicity of its main idea we include our re-implementation of SinkVC as a baseline, though the official implementation has not been published yet. 

\section{Method}

Our method follows the common ``encoder-converter-vocoder'' architecture. Similarly to kNN-VC, we use the WavLM-Large model as the encoder \cite{chen2022wavlm} and the HiFi-GAN vocoder \cite{kong2020hifi}. The conversion algorithm relies on the specific properties of the WavLM embeddings. These properties are discussed in more detail below.

\begin{figure}[t]
    \centering
    \includegraphics[width=0.95\columnwidth]{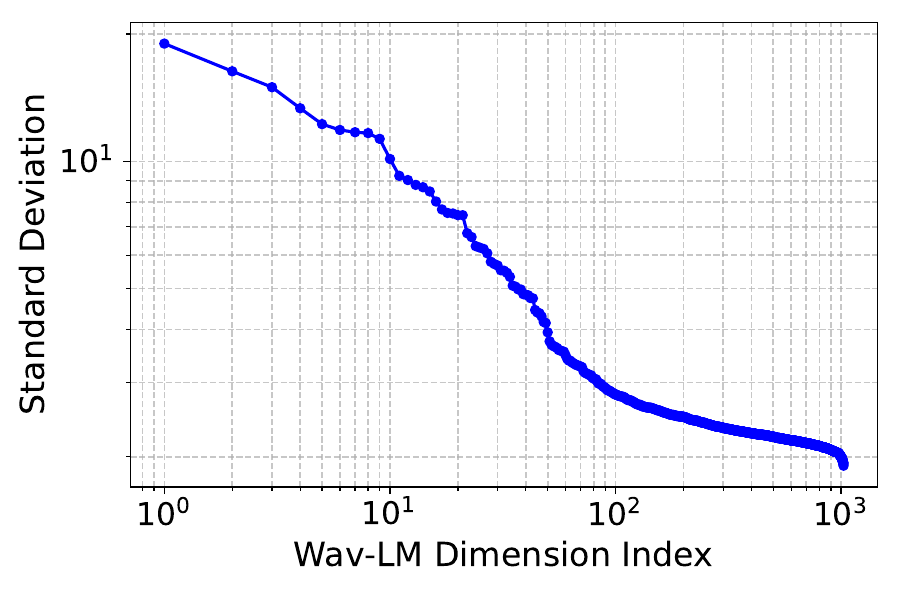}
    \caption{Sorted standard deviations of WavLM embedding components across time dimensions, plotted on a log-log scale.  Standard deviations were computed over concatenated sequences from the LibriSpeech train-clean-100 set.  Most components have small standard deviations over time and thus do not significantly contribute to pairwise L2 and cosine distances.}
    \label{fig:std_of_wavlm_embeddings}
\end{figure}

\textbf{Structure of WavLM Embeddings}
We observe that in WavLM embeddings only a small subset of dimensions has the significant numerical variability across the time axis, see Fig.\ref{fig:std_of_wavlm_embeddings}.
This implies that the L2 (or cosine) distance between two WavLM embeddings is primarily determined by the most significant 100 components, with the 900 out of 1024 components being effectively ignored.

To test this idea, we conducted a study with a trimmed kNN-VC algorithm that, during distance calculation, considers only the first $n$ components sorted by their standard deviation over time. Table \ref{tab:trimmed_knn} shows that the trimmed kNN-VC versions have the same scores compared to the baseline, and overall performance does not degrade, thus proving our hypothesis.

\begin{table}[th]
  \caption{Trimmed kNN-VC \cite{baas23_interspeech} versions, that considers only the first $n$ significant components of WavLM vector during the distance computation. LibriSpeech, test-clean \cite{panayotov2015librispeech}). Single utterance to single utterance conversion. The typical duration of an utterance is 5-10s. Scores for most $n$ do not degrade, meaning that only the most significant components contribute to the kNN-VC performance.}
  \label{tab:trimmed_knn}
  \centering
  \setlength{\tabcolsep}{3pt} 
  \begin{tabular}{l l r r r r}
    \toprule
    {kNN-VC version} & {Total Score $\downarrow$} & {WER $\downarrow$} & {CER $\downarrow$} & {SIM $\uparrow$} \\
    \midrule
$n=64$       & 0.349 & 30.097 & 17.395 & 96.939\\
$n=128$   	 & 0.354 &  30.521 & 17.698 & 97.035\\
$n=256$   	 & 0.354 &  30.562 & 17.724 & 97.109\\
$n=512$   	 & 0.357 &  30.883 & 17.753 & 97.174\\
$n=32$   	 & 0.370 &  31.822 & 18.594 & 96.808\\
$n=1024$ (orig.) 	 & 0.375 &  32.292 & 18.877 & 97.219\\
    \bottomrule
  \end{tabular}
\end{table}

In the same time, the less significant part of the embedding cannot be discarded by replacing it with random constants (zeros or average values), as it leads to degradation of reconstruction quality. Therefore, it contains valuable information.

This characteristic limits the applicability of optimal transport methods in the WavLM embedding space. Specifically, a transport plan would primarily minimize the cost associated with the first most significant dimensions, while the contribution of the remaining dimensions would not affect the transport plan. We argue that the straightforward application of the optimal transport setting as in Asadulaev et al. (2024) may be suboptimal, and the inherent structure of WavLM embeddings must be considered.

\textbf{Factorized Optimal Transport} To prevent information loss from less significant components, we propose a factorized optimal transport approach, see Fig.\ref{fig:scheme}.  This approach sorts dimensions by standard deviation, splits the embedding into lower-dimensional groups, and solves the optimal transport problem separately for each group. This allows the transport plan to perform a transformation without discarding information, as feature values within each group are similar. Furthermore, splitting high-dimensional distributions into smaller chunks makes the optimal transport computation tractable and numerically stable. We later show that the feature distribution within each group is approximately multivariate normal.

Suppose we have two Gaussian distributions, $p_{0}(x) = \mathcal{N}(x|\mu_{1}, \Sigma_{1})$ and $p_{1}(x) = \mathcal{N}(x|\mu_{2}, \Sigma_{2})$.  The Monge-Kantorovich Linear (MKL) map \cite{pitie2007linear} defines the analytical solution of quadratic optimal transport transport between them as:
\begin{equation}
    T(x) = \mu_2 + \Sigma_1^{-1/2} \left( \Sigma_1^{1/2} \Sigma_2 \Sigma_1^{1/2} \right)^{1/2} \Sigma_1^{-1/2} (x - \mu_1)
    \label{eq:mkl_map}
\end{equation}

Our factorized approach partitions the $N$-dimensional input $x$ into $N/K$ sub-vectors of dimension $K$: $x = [x^{(1)}, \ldots, x^{(N/K)}]$.  Mean vectors $\mu_1$ and $\mu_2$ are partitioned similarly. We assume covariance matrices $\Sigma_1$ and $\Sigma_2$ are approximately block-diagonal.  For each block $i$, we define a $K$-dimensional MKL transport map $T^{(i)}$ via Eq. \ref{eq:mkl_map}. The factorized map $T: \mathbb{R}^N \to \mathbb{R}^N$ is the direct product of these $K$-dimensional maps:
\begin{equation}
    T(x) = [T^{(1)}(x^{(1)}),  \ldots, T^{(N/K)}(x^{(N/K)})]
    \label{eq:factorized_transport_map}
\end{equation}
This map transports  $p_0(x)$ to  $p_1(x)$ by performing optimal transport independently within each $K$-dimensional subspace. See Fig.\ref{fig:scheme} for  illustration of Eq. \ref{eq:factorized_transport_map}. 

\begin{figure}[t]
    \centering
    \includegraphics[width=0.95\columnwidth]{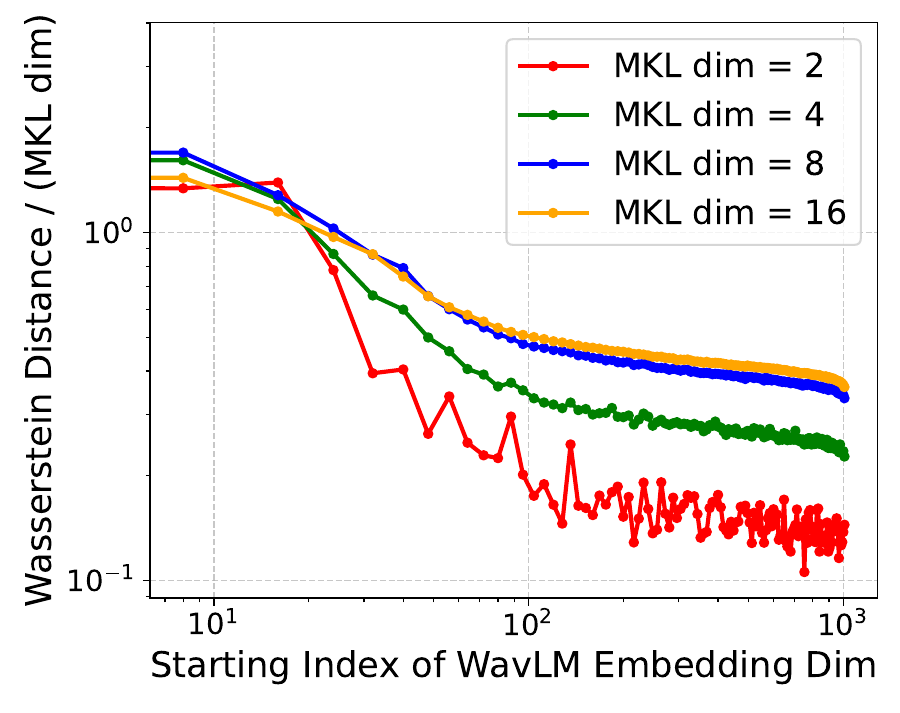}
    \caption{Wasserstein-2 distance between WavLM embeddings distribution and normal distribution with the same mean and covariance matrix, divided by the MKL dimension. LibriSpeech, train-clean-100 set.}
    \label{fig:wasserstein_distance}
\end{figure}

\textbf{Validity of Gaussian Assumption} 
Figure \ref{fig:wasserstein_distance} illustrates the Wasserstein distance between segments of WavLM embeddings (across the time axis) and a normal distribution with the same mean and covariance matrix. The x-axis represents the starting index of the WavLM embedding dimension, sampled at intervals of 8, while the y-axis depicts the Wasserstein distance. Each line in the plot corresponds to an MKL dimension $K$, ranging from 2 to 16.
By the identity of indiscernibles for the Wasserstein distance, if the  distance equals to zero, then empirical distribution is a multivariate normal.
As observed in the plot, the lower MKL dimension $K$ leads to smaller Wasserstein-2 distances to a Gaussian distribution.

We also see that Wasserstein distances in Figure 3 values are higher for lower WavLM dimensions due to larger std (as seen in Figure \ref{fig:std_of_wavlm_embeddings}). When std value is scaled by $a$, the distances are multiplied according to $d(ax, ay) = a d(x, y)$, so the Wasserstein distance also scales by the factor $a$. Therefore, the Gaussian assumption holds reasonably well across all dimensions, not just those with lower variance.

\section{Experiments}

\subsection{Datasets}

To evaluate any speaker to any speaker voice conversion we conduct our experiments on a LibriSpeech dataset \cite{panayotov2015librispeech}, which consist of 40 speakers. 
To perform the voice conversion we constructed 7800 content-reference pairs from the test-clean set of LibriSpeech \cite{panayotov2015librispeech} by taking 5 utterances for each speaker and converting them into the remaining 39 speakers, which gives a total of 40 × 5 × 39 = 7800 samples.

To check whether our method is language independent we created cross-lingual German-French dataset based on FLEURS data \cite{conneau2023fleurs}.
Both the German and French FLEURS datasets contain 10 speakers each. We generate pairs by combining every speaker from one dataset with every speaker from another, resulting in 1,000 samples.
German and French languages were selected due to presence of some sounds, such as rolling /r/ in German and its absence in French language, to see how the models are handling the generation of such specific characters.

\subsection{Baselines}

We compare the proposed approach with kNN-VC \cite{baas23_interspeech}, FACodec \cite{ju2024naturalspeech}, FreeVC and FreeVC-s\cite{li2023freevc}, Diff-VC \cite{popov2022diffusion} and our re-implementation of SinkVC (please see the Related Work section). We use for comparison the publicly available models and default settings from the official repositories. 

We compute 7800 output audios for every model tested on LibriSpeech with the exception of Diff-VC. For Diff-VC we use only 855 recordings computed in 16 hours, since its total estimated execution time on evaluation set exceeded 145 hours. 

\subsection{Metrics}

\textbf{Objective Evaluation} To evaluate speaker similarity between generated voice and target speaker reference we employ cosine similarity (SIM) between X-Vector embeddings \cite{snyder2018spoken}. To evaluate correctness of the text of the generated speech we use the Whisper-base model \cite{radford2023robust} and calculate word error rate (WER) and character error rate (CER) between the generated text and the text from the source audio. For convenience, we introduce an overall performance score, which is computed as the distance to the ideal point
\begin{equation}
    \text{Total score} = \sqrt{\text{WER}^{2} +\text{CER}^{2}+(1-\text{SIM})^{2}},
\end{equation}
where all $\text{WER}$, $\text{CER}$ and $\text{SIM}$ lie in the interval $[0,1]$ and the ideal point corresponds to $\text{WER}=0$, $\text{CER}=0$ and $\text{SIM}=1$.

\textbf{Subjective Evaluation} To evaluate the naturalness of cross-lingual conversion we have contacted people whose first languages are Arabic, Bengali, Spanish, Kazakh, Italian and Sakha (Yakut). Yakut is classified as vulnerable by the UNESCO Atlas of the World's Languages in Danger, having only 450 thousand native speakers \cite{moseley2010atlas}. 
Speech examples in these six languages were converted to a single selected reference utterance in Japanese.  
All utterances in this experiment are 10 seconds long and were taken from the Common Voice 17 dataset \footnote{\url{https://huggingface.co/datasets/mozilla-foundation/common_voice_17_0}}. We asked human experts to rank the recordings in their language from most natural to least natural. The results are given in the Table \ref{tab:human_eval}.


\begin{table}[t]
  \caption{Total scores on the test part of LibriSpeech \cite{panayotov2015librispeech} for compared baselines. Single utterance to single utterance conversion. The typical duration of an utterance is 5-10s. WER, CER and SIM reported in $\%$. Our MKL-VC algorithm is evaluated at different dimensions $K$.}
  \label{tab:aggregated_score}
  \centering
  \begin{tabular}{l r r r r}
    \toprule
    {Model} & {Total $\downarrow$} & {WER $\downarrow$} & {CER $\downarrow$} & {SIM $\uparrow$} \\
    \midrule
MKL, K=2 (ours)   	 & \textbf{0.105} &  \textbf{8.131} & \textbf{3.846} & 94.579\\
FACodec  \cite{ju2024naturalspeech} & \underline{0.106} &  \underline{8.488} & \underline{3.897} & 94.981\\
MKL, K=8 (ours)   	 & 0.111 &  8.898 & 4.344 & 94.968\\
FreeVC  \cite{li2023freevc} 	 & 0.113 &  9.061 & 4.263 & 94.795\\
MKL, K=16 (ours)   	 & 0.118 &  9.661 & 4.872 & 95.343\\
FreeVC-s  \cite{li2023freevc} 	 & 0.120 &  9.518 & 4.700 & 94.462\\
MKL, K=64 (ours)   	 & 0.151 &  12.744 & 7.213 & 96.318\\
Diff-VC$^*$  \cite{popov2022diffusion} 	 & 0.214 &  16.502 & 8.731 & 89.461\\
kNN-VC  \cite{baas23_interspeech} 	 & 0.375 &  32.292 & 18.877 & \underline{97.219}\\
SinkVC$^{**}$   	 & 0.550 &  46.891 & 28.663 & \textbf{97.691}\\
MKL, K=256 (ours)   	 & 0.827 &  65.032 & 50.775 & 93.946\\
    \midrule
    \multicolumn{5}{l}{$^*$ Diff-VC is evaluated on 855 utterances out of the 7800}\\
    \multicolumn{5}{l}{$^{**}$ SinkVC is our re-implementation with  $\varepsilon = 10^{-2}$}\\
    \bottomrule
  \end{tabular}
\end{table}

\subsection{Results}

\textbf{Comparison with Baselines} The main objective comparison with baselines is given Table \ref{tab:aggregated_score}. 
We did not concatenate or gather more than one utterance for a reference audio, so the typical duration of an utterance is 5-10s. Subjective comparison and feedback are given in Table \ref{tab:human_eval}.

MKL-VC demonstrates comparable performance to FACodec in both objective metrics and subjective evaluations, despite FACodec being trained with significantly more computational resources and MKL is training-free. The MKL-VC results show that factorized distribution matching deals with  the biggest issue of the kNN-VC pipeline, namely, its inability to handle short segments of reference speech.

Ablation study with respect to MKL dimension $K$ is included in Table \ref{tab:aggregated_score} alongside with other baselines. MKL-VC with $K=2$ reaches the highest total score due to low WER/CER and relatively good SIM. Increasing $K$ produces higher SIM scores at the cost of intelligibility. It could be explained with Fig.\ref{fig:wasserstein_distance}, where one can see that for MKL dimension $K=2$ we have the lowest Wasserstein distance, indicating that the distribution of component is closest to Gaussian.

\textbf{Cross-lingual conversion} Cross-ligual conversion study was performed on 1000 German-French pairs taken from FLEURS data \cite{conneau2023fleurs} with results presented in Table \ref{tab:cross_lang}. MKL-VC holds the second position which is in a good agreement with subjective evaluation, Table \ref{tab:human_eval}, where preferences between FAcodec and MKL-VC were divided as 50/50.



\begin{table}[th]
  \caption{Cross-lingual conversion computed on 1000 German-French pairs taken from FLEURS data \cite{conneau2023fleurs}. The typical duration of an utterance around 10s.}
  \label{tab:cross_lang}
  \centering
  \begin{tabular}{ l  r r r r}
    \toprule
    \multicolumn{1}{c}{dim $K$} & {Total $\downarrow$} & {WER $\downarrow$} & {CER $\downarrow$} & {SIM $\uparrow$} \\
    \midrule
FACodec  \cite{ju2024naturalspeech} 	 & \textbf{0.400} &  \textbf{36.659} & \textbf{15.354} & \underline{95.759}\\
MKL, K=2 (ours)   	 & \underline{0.441} &  \underline{39.168} & 19.696 & 94.793\\
    FreeVC-s  \cite{li2023freevc} 	 & 0.481 &  43.688 & \underline{19.449} & 94.598\\
kNN-VC  \cite{baas23_interspeech} 	 & 1.156 &  96.662 & 63.340 & \textbf{96.520}\\
    \bottomrule
  \end{tabular}
\end{table}

\section{Conclusion}

In our work, we propose MKL-VC, a training-free modification of the kNN-VC pipeline for any-to-any cross-lingual voice conversion. MKL stands for Monge-Kantorovich Linear mapping and is based on the exact solution of the quadratic optimal transport problem for Gaussian distributions. We show that MKL-VC solves the main issue of the original algorithm, that is, unstable performance with short reference audio.

In extensive evaluation MKL-VC demonstrates significant improvement of content-related metrics and achieves performance comparable to FACodec \cite{ju2024naturalspeech} handling only 5-10 seconds of reference audio instead of 5 minutes required for kNN-VC. This is due to MKL-VC matches distributions and can produce new tokens which are absent in a target speaker embeddings.

However, the proposed scheme is not universal because it relies on specific properties of WavLM embeddings.  For a new encoder, the Gaussianity assumptions must be explicitly verified, as they are not guaranteed to hold in general.

Like its predecessor, MKL-VC has only one tuning parameter, $K$, which governs the trade-off between content and reference similarity: a higher MKL dimension $K$ corresponds to higher speaker similarity but lower content scores.

According to the feedback received, the distinguishing feature of the MKL-VC approach is the absence of a robotic voice effect. This method can help generate movie scoring in any low-resource language, such as Sakha, which does not have large corpora of datasets to train on, while still maintaining good FACodec-level quality.



\begin{table}[t]
  \caption{Human evaluation and feedback. Cross-lingual conversion with reference speech in Japanese and source speech in six languages. Six native speakers were asked to rank four audio examples from the most natural to the least natural. Numbers were used for anonymization: 1 is Diff-VC, 2 is FACodec, 3 is kNN-VC, 4 is MKL-VC (ours).}
  \label{tab:human_eval}
  \centering
  \begin{tabular}{l r p{3.0cm}}
    \toprule
    {Human expert} & Lang.& {Ranking} \\
    \midrule
    Arabic      &[ar]  & 2 4 3 1 \\
    Bengali (Bangladesh)    &[bn]  & (2 and 4) then (3 and 1) \\
    Spanish (México) &[es]  & 2 4 3 1 \\
    Kazakh      &[kk]  & 4 2 3 1 \\
    Italian     &[it]  & 4 then (2 and 3) then 1 \\
    Sakha       &[sah] & 4 2 3 1 \\
    \midrule
  \end{tabular}
  \begin{tabular}{l p{6.5cm}}
    \multicolumn{2}{l}{Comments:}\\
    {[bn]}  & Only 2 and 4 are close to Bengali, they sounds almost identical. \\
    {[es]} &  2 sounds the most natural and mexican, although ``s'' sounds are a bit muted,
    4 is good and clear but the intonation is slightly worse than the other, 3 sounds like an old Spanish person. 1 is having a stroke \\
    {[sah]} & 4 is more natural, 2 has distortions (robotic voice), 3 and 1 are like uncanny valley.\\
    \bottomrule
  \end{tabular}
\end{table}

\section{Broader Impact Statement}
Since VC models can synthesize speech with high speaker similarity, they carry potential risks, such as spoofing voice identification systems or impersonating specific speakers. To prevent misuse, it is crucial to develop robust speech detection methods and avoid voice authentication in high-security applications.

\bibliographystyle{IEEEtran}
\bibliography{mybib}

\end{document}